\documentclass[dvips]{article}

\usepackage{icrc2011}

\title{\HESS\ deeper observations on SNR \velajr}
\shorttitle{M. Paz Arribas \etal\ SNR \velajr\ with \HESS}

\newcommand{\velajr}{RX~J0852.0$-$4622}
\newcommand{\velajrg}{G266.2$-$1.2}
\newcommand{\psr}{PSR~J0855$-$4644}
\newcommand{\rxj}{RX~J1713.7$-$3946}
\newcommand{\HESS}{H.E.S.S.}

\newcommand{\ATNF}{ATNF}
\newcommand{\CANGAROO}{CANGAROO}
\newcommand{\ROSAT}{\emph{ROSAT}}
\newcommand{\ASCA}{\emph{ASCA}}

\newcommand{\FERMI}{\emph{Fermi}}

\newcommand{\s}{\mathrm{s}}
\newcommand{\ms}{\mathrm{ms}}
\newcommand{\cm}{\mathrm{cm}}
\newcommand{\kpc}{\mathrm{kpc}}
\newcommand{\erg}{\mathrm{erg}}
\newcommand{\TeV}{\mathrm{TeV}}
\newcommand{\GeV}{\mathrm{GeV}}
\newcommand{\etal}{\MakeLowercase{\textit{et al.}}} 

\authors{Manuel Paz Arribas$^{1,2\,*}$, Ullrich Schwanke$^{2}$, Iurii Sushch$^{2,8}$, Nukri Komin$^{3}$, Fabio Acero$^{4}$ and Stefan Ohm$^{5,6,7}$ for the \HESS\ Collaboration}
\afiliations{
  $^1$DESY, D-15735 Zeuthen, Germany\\
  $^2$Institut f\"ur Physik, Humboldt-Universit\"at zu Berlin, Newtonstr. 15, D 12489 Berlin, Germany\\
  $^3$LAPP, Universit\'{e} de Savoie, CNRS/IN2P3, F-74941 Annecy-le-Vieux, France\\
  $^4$LUPM, Universit\'{e} Montpellier II, CNRS/IN2P3, F-34095 Montpellier, France\\
  $^5$Max-Planck-Institut f\"ur Kernphysik, P.O. Box 103980, D 69029 Heidelberg, Germany\\
  $^6$School of Physics \& Astronomy, University of Leeds, Leeds LS2 9JT, UK\\
  $^7$Department of Physics and Astronomy, The University of Leicester, Leicester LE1 7RH, UK\\
  $^8$Taras Shevchenko National University of Kyiv, Ukraine
}
\email{* mapaz@physik.hu-berlin.de}

\abstract{Supernova Remnants (SNRs) are believed to be acceleration sites of
Galactic cosmic rays. Therefore, deep studies of these objects
are instrumental for an understanding of the high energy processes in
our Galaxy.
\velajr, also known as Vela Junior, is one of the few (4)
shell-type SNRs resolved at Very High Energies (VHE; $E > 100\,\GeV$). It
is one of the largest known VHE sources ($\sim 1.0^\circ$ radius) and its flux level is
comparable to the flux level of the Crab Nebula in the same energy band.
These characteristics allow for a detailed analysis, shedding further
light on the high-energy processes taking place in the remnant.
In this document we present further details on the spatial and spectral
morphology derived with an extended data set.
The analysis of the spectral morphology of the remnant is
compatible with a constant power-law photon index of $2.11 \pm 0.05_{\rm stat} \pm 0.20_{\rm syst}$ from the whole SNR in the energy range from 0.5 TeV to 7 TeV.
The analysis of the spatial morphology shows an enhanced emission towards
the direction of the pulsar \psr, however as the pulsar is lying on the rim of the SNR,
it is difficult to disentangle both contributions. Therefore, assuming a point source,
the upper limit on the flux of the pulsar wind nebula (PWN) between 1 TeV and 10 TeV,
is estimated to be $\sim 2\%$ of the Crab Nebula flux in the same energy range.}

\begin{document}
\maketitle

\section{Introduction}

\velajr\ is a supernova remnant (SNR) located towards the direction of the constellation Vela. Its projection on the sky overlaps with the wider Vela SNR, for this reason, it is commonly known as ``Vela Junior''.

\velajr\ was first discovered in the late 90s during the all-sky survey performed by \ROSAT\ at X-ray energies \cite{paper:rosat}. Since then, it has been observed in a variety of wavelengths, ranging from radio to VHE $\gamma-$rays. It has been spatially resolved as a shell-type SNR, with an apparent size of $2^\circ$ in diameter.

Further details like the type of the SN explosion that originated Vela Junior, its age and distance are subject to controversy, due to the complexity of the region (see \cite{paper:mwl} and reference therein for more details).

In $\gamma-$rays, emission coming from the north-western part of the rim was first detected by \CANGAROO\ \cite{paper:cangaroo} at the VHE regime. Further observations with the \HESS\ experiment \cite{paper:hess} have shown a $2^\circ$ extended emission covering the whole area of the SNR, with a spatially resolved shell morphology. This makes Vela Junior one of the largest objects in the VHE sky. The spectrum of the emission was well described by a simple power law.

The \FERMI\ collaboration has recently reported the analysis of \velajr\ \cite{talk:fermi}. An extended emission at the nominal position of the SNR with roughly the same size as the \HESS\ source was detected. Although the shell morphology could not be resolved, the emission was well described by a power-law spectrum, in the energy range of a few GeV up to a few hundred GeV, which connects well with the \HESS\ points.

Recent X-ray observations \cite{proc:psr_crism} revealed a pulsar wind nebula (PWN) on the rim of \velajr\ around the energetic radio pulsar \psr. The properties of the pulsar are listed in tab. \ref{tab:psr}. The estimate of the distance of 4 kpc shown in the table is conservatively high. It was derived from the dispersion measure (DM) using the model by Cordes \& Lazio from 2002 \cite{paper:dm_model}. More recent studies, based on column density measurements in X-rays, indicate that the pulsar could be closer than 0.9 kpc \cite{proc:psr_crism}. For this reason, the DM estimate is taken as a conservative upper limit of the distance.

\begin{table}[!htb]
  \begin{center}
    \begin{tabular}{rcl}
      period              & $=$ & $64\,\ms$                      \\
      characteristic age  & $=$ & $140\,\mathrm{kyr}$            \\
      distance            & $<$ & $4\,\kpc$                      \\
      spin-down power     & $=$ & $1.1 \times 10^{36}\,\erg/\s$  \\
    \end{tabular}
    \caption{Properties of \psr. The information was taken from the \ATNF\ pulsar catalogue \cite{atnf_psr_catalogue}.}\label{tab:psr}
  \end{center}
\end{table}

Since the last \HESS\ publication, the amount of data available for \velajr\ has increased by a factor of two. In the following, we present the results of the analysis of the spectral and spatial morphology of the complete data set.

\section{\HESS\ observations}\label{sec:hess}

\HESS\ is an array of four Cherenkov telescopes situated in the Khomas Highland of Namibia at an altitude of 1800 m above sea level. It is sensitive in the energy range of 100 GeV to 100 TeV and can detect a point source with a flux level of 1\% of the Crab Nebula flux in the same energy range in 25 hours at low zenith angles. Its wide field of view ($5^\circ$ diameter), angular resolution ($0.1^\circ$ per event) and energy resolution (15\% to 20\% per event) make it very suitable for spectral and morphological analysis of very extended sources like \velajr.

The data used for the analysis were taken between 2004 and 2009. The events were reconstructed using a Hillas parameter technique \cite{paper:reco} before background subtraction. Statistical significances were calculated with the Li\&Ma formula \cite{paper:lima}.\footnote{Software tag: \textbf{hap-11-02-pl07}.} The results where crosschecked using an independent analysis and calibration method, yielding consistent results.

The data available after data quality selection and dead time correction, amounts to 7286 photons within 72 h of livetime, resulting in a significance of $37\sigma$.

\begin{figure}[!htb]
  \centering
  \includegraphics[width=3in]{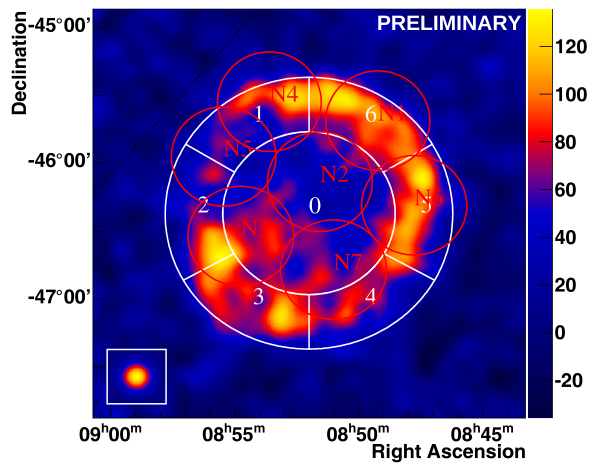}
  \caption{Exposure-corrected excess map smoothed with a gaussian function with a standard deviation of $0.06^\circ$. The white lines represent the different regions selected for spectrum extraction. The red lines indicate the different pointing observations of the \ASCA\ satellite. The black dashed line represents the Galactic Plane. The inset represents the PSF of the instrument for this particular analysis.}
  \label{fig:skymap}
\end{figure}

In the case of the spatial morphology analysis, a hard selection of cuts was applied to the data. These cuts improve the angular resolution at the expense of reduced statistics and a higher energy threshold. In order to model the background, a ring model was used \cite{paper:bg}. The resulting photon excess corrected for the different exposure in each point of the sky, is shown in fig. \ref{fig:skymap} smoothed with a gaussian function with a standard deviation of $0.06^\circ$.

\subsection{Spectral morphology}

With the current statistics, a study of the spectral morphology is possible. In order to perform this study, the SNR was divided into several regions, as shown in white in fig. \ref{fig:skymap}: a central circle (region 0) and 6 segments of a shell following the shape of the emission (regions 1 to 6).

A particular run selection was performed for each of the regions. In this case, standard selection cuts were applied in order to have enough statistics, and a reflected region model was used for background modeling \cite{paper:bg}. This method is more suited for flux calculations since it has less systematic effects.

\begin{figure*}[!htb]
  \centerline{
    \includegraphics[width=3in]{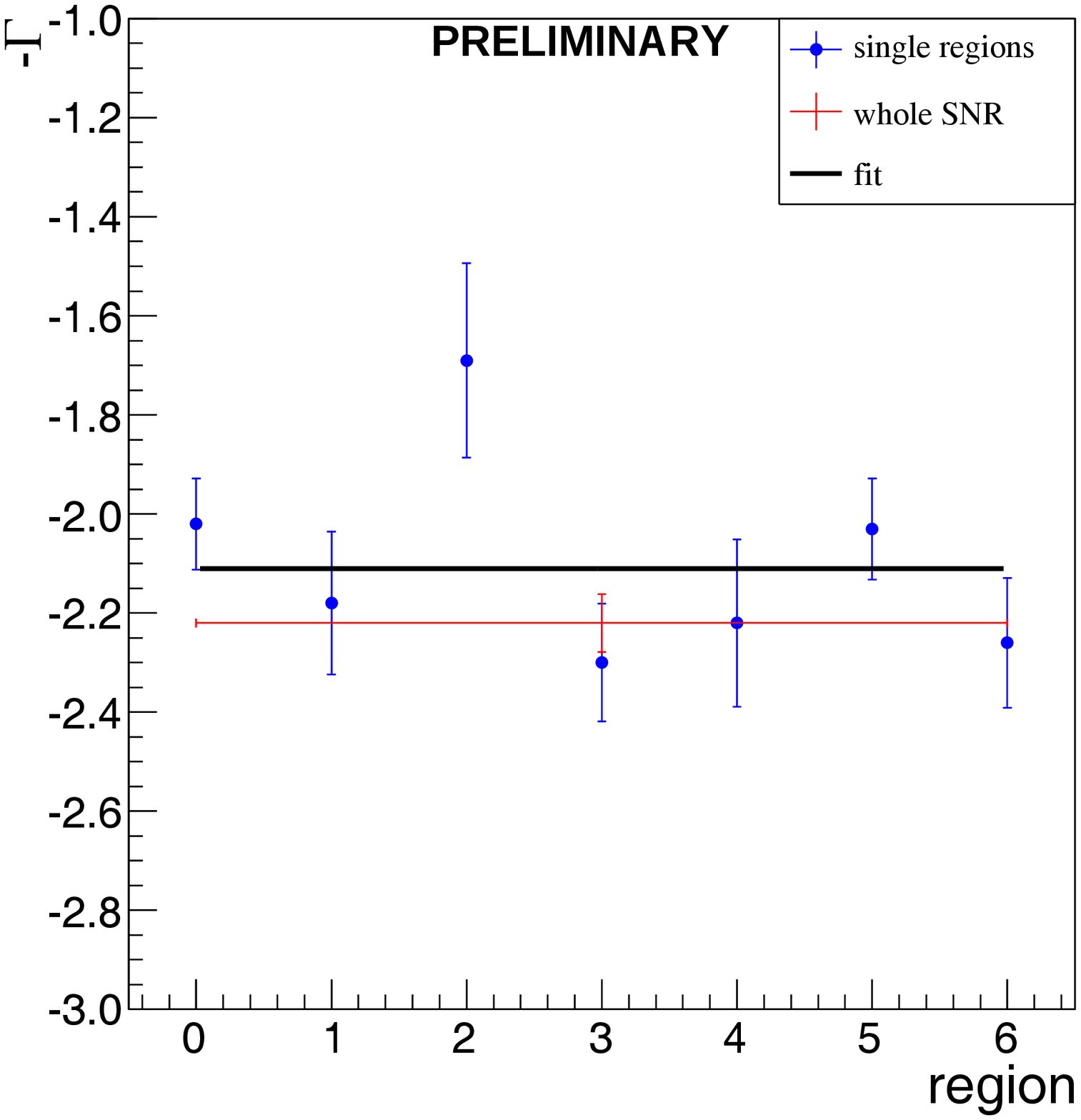}
    \hfil
    \includegraphics[width=3in]{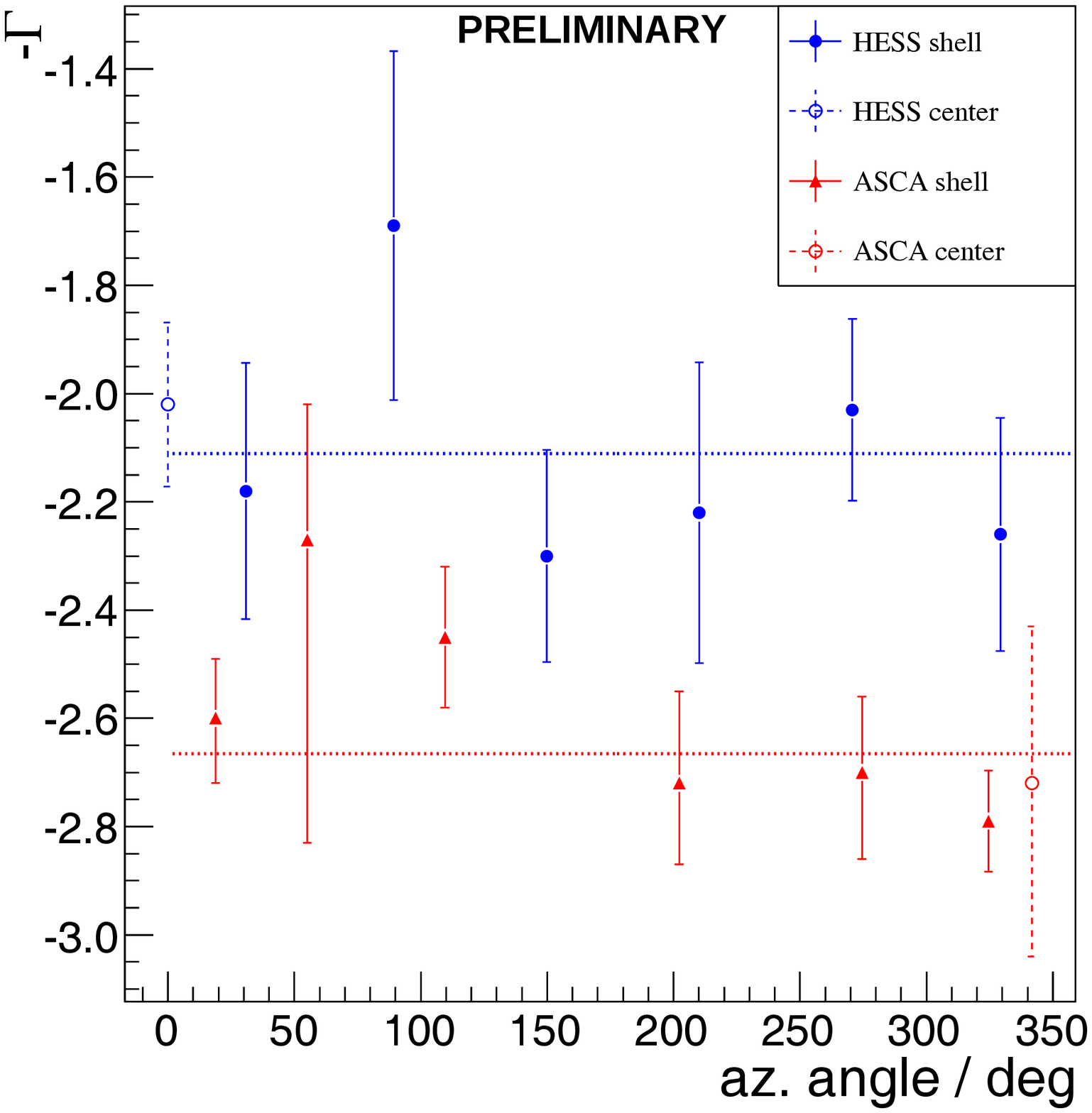}
  }
  \caption{Spectral morphology of the remnant. Left pad: plot of the spectral index of each of the \HESS\ regions denoted in white in fig. \ref{fig:skymap} (blue points). The red point represents the same measurement, but for the whole SNR. The black line represents the fit to the blue points. Right pad: plot of the spectral indices for the \HESS\ (blue points) and \ASCA\ (red points) regions shown in fig.\ref{fig:skymap}, as a function of the azimuthal angle of the centroid of the regions. The errors are given at the 90\% confidence level (C.L.). The horizontal lines represent the constant values fitted to the \HESS\ and \ASCA\ points. The definition of the azimuthal angle is shown by the red arrow in fig. \ref{fig:az_prof} left pad.}
  \label{fig:spec_morph}
\end{figure*}

The spectrum derived for each of the regions was fitted with a power-law in the range from 0.5 TeV to 7 TeV. In order to test for changes in the spectral slope, the indices for all regions were fitted to a constant function, yielding a value of $2.11 \pm 0.05_{\rm stat} \pm 0.20_{\rm syst}$, with a $\chi^2$ of 10.7 for 6 degrees of freedom (fit probability of $9.9\%$). The different indices together with the fit result and the index obtained for the whole SNR ($2.22 \pm 0.06_{\rm stat} \pm 0.20_{\rm syst}$) are shown in the left pad of fig. \ref{fig:spec_morph}. The figure shows that the indices of all the regions are compatible with the constant value; only region 2 seems to have a harder spectrum, which is still compatible with the constant within $2.1\sigma$.

A similar measurement of the spectral morphology was performed in X-rays by \ASCA. Since the field of view of the satellite is smaller than the angular size of \velajr, a series of 7 pointings was performed with the GIS camera, which is sensitive in the 0.7 keV to 10 keV energy band, in order to cover most of the SNR \cite{paper:asca}. The \ASCA\ pointings are shown in red in fig. \ref{fig:skymap}. The indices for the spectra derived for the different pointings of the \ASCA\ satellite \footnote{\ASCA\ spectral analysis results provided by Junko S. Hiraga.}, together with the \HESS\ indices, are shown as a function of the azimuthal angle in the right pad of fig. \ref{fig:spec_morph}. Like for the \HESS\ data, a constant function was fitted to the \ASCA\ spectral indices, yielding a value of $2.66 \pm 0.05_{\rm stat}$ (error at the 90\% C.L.).

\subsection{Spatial morphology}

In the beginning of section \ref{sec:hess}, the data available for spatial morphology analysis was presented. This data was used to produce the skymap in fig. \ref{fig:skymap}, that shows that the emission comes mostly from a thin ($\sim 0.2^\circ$) shell on the border of the SNR, in agreement with the radial profile from \cite{paper:hess}.

\begin{figure*}[!htb]
  \centering
  \includegraphics[width=7in]{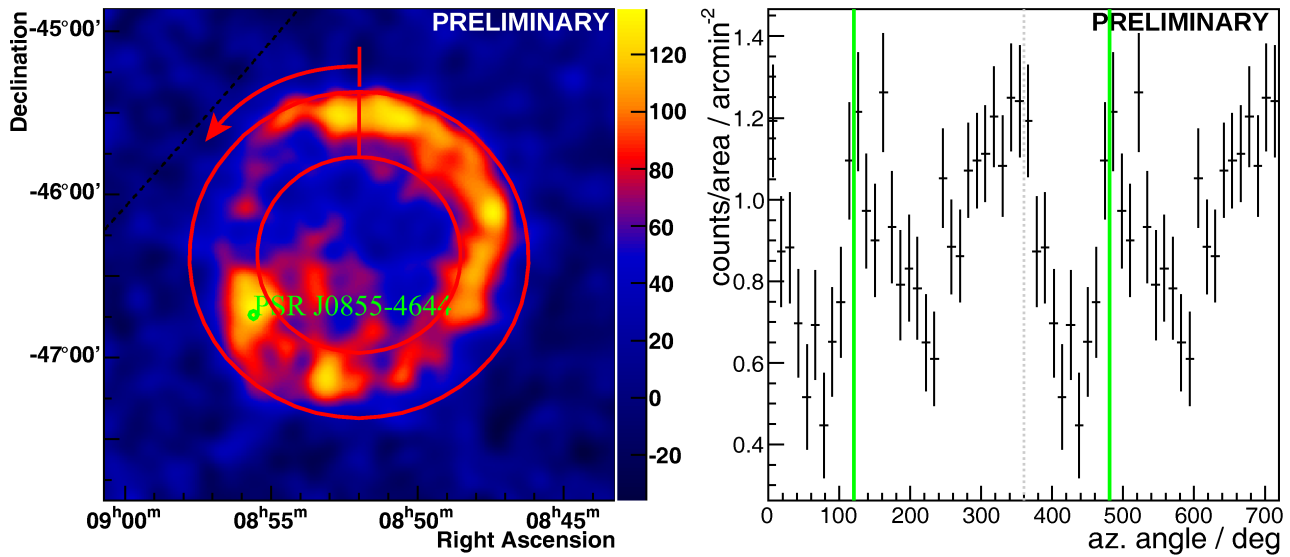}
  \caption{Azimuthal profile from the shell of \velajr. The red region on the skymap (left pad) indicates the extraction region for the profile (right pad): for clarification two periods are shown, separated by a gray dotted line. The definition of the azimuthal angle is shown by the red arrow in the skymap. The position of \psr\ is marked with a green circle in the skymap, and its projection in the profile extraction region is denoted with a green line in the histogram. The black dashed line in the skymap represents the Galactic Plane.}
  \label{fig:az_prof}
\end{figure*}

In order to look for structures in the shell, an azimuthal profile was derived from the unsmoothed exposure-corrected excess map for an annular region centered at the nominal position of the SNR between $0.6^\circ$ and $1.0^\circ$, as shown on the left pad of fig. \ref{fig:az_prof}. The profile itself is shown in the right pad of fig. \ref{fig:az_prof}. The values in the histogram are corrected for the area of the corresponding region in the sky.

The azimuthal profile shows that the emission is not homogeneous along the shell. The north-western rim (from $270^\circ$ to $380^\circ$ approx.) is brighter than the south-eastern part, with some local enhancements, like the one seen at about $120^\circ$ towards the direction of \psr. The projection of the pulsar position in the profile extraction region is marked with a green line in the histogram.

This morphology is also seen in radio and X-ray maps, where most of the emission comes from the north-western parts of the shell, with some local enhancements towards the south. In the case of the X-rays an enhancement is also seen towards the direction of \psr.

\subsection{Upper limit on \psr}

According to the pulsar population study published in \cite{proc:psr_pop_study}, pulsars with $\dot{E}/d^2 > 10^{34}\,\erg/\s/\kpc^2$ are very likely of being detected as PWNe in $\gamma-$rays. In the case of \psr, $\dot{E}/d^2 > 6.9 \times 10^{34}\,\erg/\s/\kpc^2$, which corresponds to a detection probability $> 20\%$. Using the more recent estimate for the distance of 0.9 kpc $\dot{E}/d^2 = 1.4 \times 10^{36}\,\erg/\s/\kpc^2$, for which the detection probability in $\gamma-$rays amounts to $> 70\%$.

Since the pulsar emission overlaps with the emission of the shell of \velajr\ in $\gamma-$rays, it is not possible to disentangle, with the current angular resolution, the possible PWN contribution from the shell of the SNR. Assuming a point source centered at the pulsar position, a spectral fit was performed on the region, from which the integral flux was calculated and interpreted as an upper limit on the flux of the PWN in $\gamma-$rays:\newline
$F(1\,\TeV < E < 10\,\TeV) < 3.8 \times 10^{-13}\,\cm^{-2}\,\s^{-1}$.\newline
This flux represents $\sim 2\%$ of the Crab Nebula flux in the same energy range.

\section{Conclusions}

The \HESS\ analysis of an extended data set on the Vela Junior region in the VHE regime has been presented. The increased statistics permitted a more detailed study of the morphology of the SNR.

In the case of the spectral morphology, a constant index is found across the whole SNR in the energy range from 0.5 TeV to 7 TeV, compatible with the value of $2.11 \pm 0.05_{\rm stat} \pm 0.20_{\rm syst}$. This value is found to differ by 0.55 with the one derived for X-rays ($2.66 \pm 0.05_{\rm stat}$ at the 90\% C.L.). In the framework of a very simple leptonic model, in which the same population of electrons with a power-law electron energy distribution in a homogeneous magnetic field would be responsible for both the synchrotron and the inverse Compton emissions, the same photon index is expected for both components. The difference observed here could be explained by cooling of the high energy electrons, who are responsible for the synchrotron emission in X-rays. On the other hand, this is not a very strong argument, and indeed a hadronic scenario cannot be ruled out. As for the case of \rxj\ more detailed spectra of the whole SNR in $\gamma-$rays with \HESS\ and \FERMI\ would be very useful for revealing the nature of the emission.

Regarding the spatial morphology of the remnant, an enhancement is seen towards the direction of \psr. Since the contribution from the PWN cannot be disentangled from the emission from the shell of \velajr, an upper limit on the flux of a possible VHE PWN between 1 TeV and 10 TeV, assuming a point source is derived at the level of $\sim 2\%$ of the Crab Nebula flux in the same energy range.

\clearpage

\end{document}